\documentclass[prl,twocolumn,showpacs]{revtex4}
\usepackage{graphicx}
\usepackage{dcolumn}
\usepackage{amssymb}
\usepackage{bm} 
\def\Q{\mbox{\sffamily\bfseries Q}}
\def\T{\mbox{\sffamily\bfseries T}}

\def\I{\mbox{\sffamily\bfseries I}}

\def\G{\mbox{\sffamily\bfseries G}}

\def\nabbold{\mbox{\boldmath $\nabla$\unboldmath}}
\def\otens{\mbox{\boldmath $\Omega$\unboldmath}}

\def\kapbold{\mbox{\boldmath $\kappa$\unboldmath}}

\def\sigop{\mbox{\boldmath $\sigma$\unboldmath$^{\tiny OP}$}}

\def\beq{\begin{equation}}                           
\def\eeq{\end{equation}}                           
\def\bea{\begin{eqnarray}}                           
\def\eea{\end{eqnarray}}                           
\begin{document}
\title{Spatiotemporal rheochaos in nematic hydrodynamics} 
\author{Buddhapriya Chakrabarti$^1$\cite{a},} 
\email{buddho@physics.umass.edu}
\author{Moumita Das,} 
\email{moumita@physics.iisc.ernet.in} 
\author{Chandan Dasgupta$^1$,} 
\author{Sriram Ramaswamy$^1$,} 
\author{A.K. Sood}
\email{asood@physics.iisc.ernet.in}
\affiliation{Department of Physics,
Indian Institute of Science, Bangalore 560012 INDIA\\
$^1$Condensed Matter Theory Unit, JNCASR, Bangalore 560064 INDIA}
\begin{abstract} 
Motivated by recent observations of rheochaos in sheared wormlike micelles,  
we study the coupled nonlinear partial differential equations for 
the hydrodynamic velocity and order parameter fields in a sheared 
nematogenic fluid.  
In a suitable parameter range, we find irregular, dynamic shear-banding 
and establish by decisive numerical tests that the chaos we observe 
in the model is spatiotemporal in nature.   
\end{abstract}

\maketitle

Wormlike micellar solutions of 
cetyltrimethylammonium tosylate (CTAT) show rheochaos 
\cite{aks1,aks2,aks3}: for a suitable range of concentrations and temperatures, 
the stress {\it vs.} shear-rate curve in \cite{aks1,aks3} 
shows a plateau signalling a rheological phase-coexistence  
and, in the plateau region, the time-series of the shear and normal stresses 
at constant shear-rate show deterministic chaos. 
This is remarkable because it means that the rheological 
equation of state \cite{larsonbook}, that is, the relation between stress and shear-rate, 
even for macroscopic experimental samples, has huge fluctuations about the mean 
\cite{ramamohan}. 
A realistic description of the complex dynamics of micellar systems 
as well as of other rheological oscillations \cite{hlpap1,roux} may require  
variables specific to the system in question \cite{headcates1,catesolmsted03}. 
The  studies of \cite{grosso,rien1,rien2}, however, are significant in that they   
find temporal rheochaos in the dynamics of the passively advected alignment tensor alone. 
They use the well-established equations of hydrodynamics for a nematic order 
parameter, outside the regime of stable flow alignment, but consider 
only spatially {\em homogeneous} states \cite{znat} and hence cannot explore the relation 
of the observed chaos to shear-banding \cite{band}. 

\begin{figure}
\includegraphics[width=8.5cm,height=5.2cm]{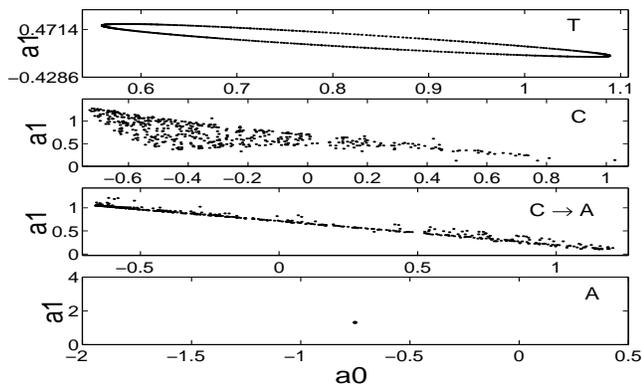}
\caption{\label{phport}Plots showing $a_1(x_0,t)$ vs $a_0(x_0,t)$ for 
tumbling (panel T), chaotic (C), 
onset of aligning (C$\to$A) and aligning (A) regimes.}
\end{figure}

\begin{figure}
\includegraphics[width=8.5cm,height=5.2cm]{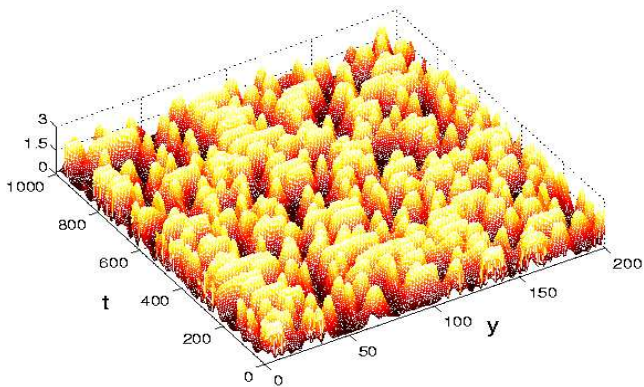}
\caption{\label{ChShearstress}Space-time behavior of the shear stress in the chaotic regime, 
$\dot{\gamma}=3.678$ and $\lambda_k$=1.25. Slice taken from a system of 
size 5000}
\end{figure}

\begin{figure}
\includegraphics[width=8.5cm,height=5.2cm]{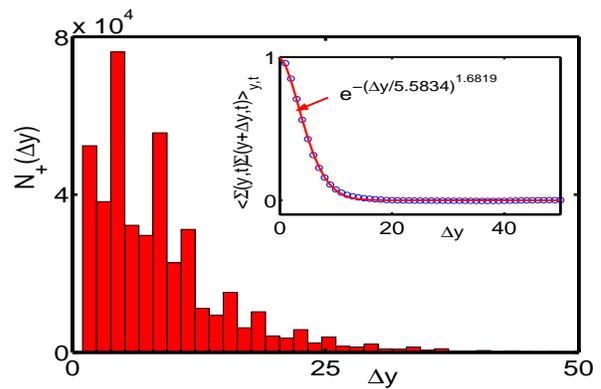}
\caption{\label{banddistrib} Spatial distribution of ``stress drops'' (corresponding
to residence intervals in which the shear stress is above a threshold $\Sigma_0=0.8$) in
the Chaotic regime (Fig. 3).
Inset shows the spatial autocorrelation function averaged over t=5000 (circles)
with the best fit to the form $\exp[(-\Delta y/\xi)^{\alpha}]$.}  
\end{figure}

\begin{figure}
\includegraphics[width=8.5cm,height=5.2cm]{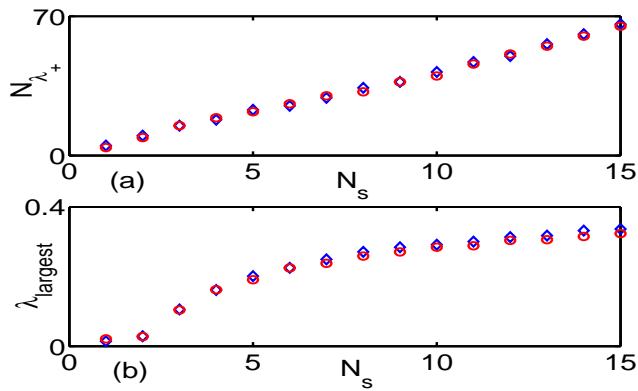}
\caption{\label{Npositivelambda}\small Number of positive
Lyapunov exponents (top panel) and the largest Lyapunov exponent (bottom
panel) as functions of subsystem size $N_s$, for $\dot{\gamma}=3.678$, 
$\lambda_k=1.25$, 
Embedding dimension for
the time series of each space point is 10. Diamonds and 
circles correspond to data taken from two different regions of space
($i_0$=51 and $i_0$=101 respectively, see text).} 
\end{figure}

In this paper we study numerically the equations 
of the traceless symmetric order 
parameter for a sheared nematogenic system, allowing for spatial 
variation. In the parameter range termed 
``Complex'' in the phase diagram of the studies 
\cite{rien1,rien2} of the spatially homogeneous situation, 
we find spatiotemporal chaos.  
We give evidence for this in Fig. \ref{ChShearstress} for the dynamic 
instability of shear bands,  
Fig. \ref{banddistrib} for the spatial distribution of ``stress drops'', 
Fig. \ref{phport}(b)
for the local phase portrait in the chaotic regime, 
and, decisively, Fig. \ref{Npositivelambda} 
for the number of positive Lyapunov exponents as 
a function of (sub)system size. 
We now present our model in detail and show how these results arise. 

Models of complex rheology \cite{larsonbook} generally postulate an equation 
of motion for the stress in addition to that for the velocity field. 
Our work was motivated in part by the failure \cite{buddhothesis} 
to find chaos in the 
the popular Johnson-Segalman (JS) \cite{johnsegman1} model 
for the coupled dynamics of the stress tensor and the 
hydrodynamic velocity field. Since this model 
is {\em linear} in the stress, and contains bilinear 
products of velocity-gradient and stress,  
it is natural to ask  
if nonlinearities in the stress are the missing ingredient. 
The JS equation is most naturally 
thought of as arising from the 
underlying dynamics of a local alignment tensor, 
the traceless, symmetric nematic order parameter $\Q$, 
measuring in this case the alignment of the micellar worms.  
The stress $\sigop$ \cite{deviatoric} due to $\Q$ is defined 
via a free-energy functional $F[\Q]$ \cite{rien2,degp,forster,olmstedgoldbart}. 
Nonlinearities arise naturally 
in nematodynamics from terms of order higher than $\Q^2$ in $F$. 
It is these that lead to the rich dynamical phase diagram, including a 
variety of routes to chaos, in the work of \cite{grosso,rien1,rien2}.  

The alignment tensor obeys the equation of motion 
\bea 
\label{rqeq1}
{\partial \Q \over \partial t}
+{\bf u}\cdot \nabbold \Q &=&   
 \tau^{-1} \G  
            +  (\alpha_0 \kapbold + \alpha_1 \kapbold \cdot \Q)_{ST} \nonumber\\
&+& 
\otens \cdot \Q - \Q \cdot \otens, 
\eea
where $\I$ is the unit tensor, ${\bf u}$ is the hydrodynamic velocity field, 
$\kapbold \equiv (1/2)[\nabbold {\bf u} + (\nabbold {\bf u})^T]$ 
and $\otens \equiv (1/2)[\nabbold {\bf u} - (\nabbold {\bf u})^T]$
the shear-rate and vorticity tensors respectively, $\tau$ is a bare relaxation 
time, $\alpha_0$ and $\alpha_1$ are parameters related to flow alignment, 
the subscript $ST$ denotes symmetrisation and trace-removal, and  
\bea 
\label{molfield} 
\G \equiv 
-(\delta F / \delta \Q)_{ST} 
&=&  -[A \Q - \sqrt{6} B (\Q \cdot \Q)_{ST} + C \Q \Q : \Q \nonumber \\
&+& \Gamma_1 \nabla^2 \Q + 
\Gamma_2 (\nabbold \nabbold \cdot \Q)_{ST}] 
\eea
is the molecular field conjugate to $\Q$, 
for a Landau-de Gennes \cite{degp} 
free-energy functional $F$ with upto quartic terms in $\Q$, 
with phenomenological coefficients $A$, $B$, $C$, $\Gamma_1$, $\Gamma_2$,  
and the simplest 
bilinears in $\nabbold \Q$.  
In mean-field theory, 
the isotropic-nematic transition occurs at $A = A_* = 2B^2/9C$.   
As in \cite{grosso,rien1,rien2}, we 
impose plane Couette flow with $\mathbf{u} = y \dot{\gamma} \hat{x}$, 
ignoring modifications of this flow by order-parameter stresses, 
and treating $\dot{\gamma}$ as a control parameter, 
but, {\em unlike} in that work, we allow spatial inhomogeneity in 
the $y$ direction.  
 
As in \cite{rien2}, we rescale time by the relaxation time $\tau/A_*$ 
at the isotropic-nematic transition, $\Q$ as well by its magnitude at 
that transition, and distances by 
the diffusion length made from $\Gamma_1$ and $\tau/A_*$.
The ratio
$\Gamma_2 /\Gamma_1$ of the Frank constants is therefore a free parameter which 
we have set to 1 in our study. 
We choose $A = 0$ and $\alpha_1 = 0$ throughout, 
and $\lambda_k = -(2/\sqrt{3})\alpha_0$ and $\dot{\gamma}$ 
with values as specified in the figures. 
In the orthonormal basis \cite{rien1,rien2}  
$\T_0 = \sqrt{3/2}(\hat{\bf z} \hat{\bf z})_{ST}$, 
$\T_1 = \sqrt{1/2}(\hat{\bf x} \hat{\bf x} - \hat{\bf y} \hat{\bf y})$, 
$\T_2 = \sqrt{2}(\hat{\bf x} \hat{\bf y})_{ST}$, 
$\T_3 = \sqrt{2}(\hat{\bf x} \hat{\bf z})_{ST}$, 
$\T_4 = \sqrt{2}(\hat{\bf y} \hat{\bf z})_{ST}$, 
we expand $\Q = \sum_i a_i \T_i$ and work with the equations of motion for the 
five components $a_i$ which follow from (\ref{rqeq1}).   
These equations are numerically integrated using a 4th order Runge-Kutta scheme with a fixed 
time step ($\Delta t = 0.001$). For all the results quoted here a symmetrized form 
of the finite difference scheme involving nearest neighbors is used to calculate the 
gradient terms. We have checked that our results are not changed if smaller values of 
$\Delta t$ are used. We have further checked that the results do not change if the 
grid spacing is changed and more neighbors to the left and right of a particular site 
in question are used to calculate the derivative. This gives us confidence that the 
results quoted here do reflect the behavior of a continuum theory and are not artifacts 
of the numerical procedure used. We use boundary conditions with the director being normal to the walls. With this, we discard the 1st $6 \times 10^6$ timesteps
to avoid any possibly transient behavior. We monitor the time evolution of the system 
for the next $5 \times 10^6$ time steps (i.e. t=5000), recording configurations after every 
$10^3$ steps.
We have carried out the study with system sizes ranging from 100 to 5000. 
In the time-series analysis for the Lyapunov spectrum, we run the simulation
till $t=20,000$, for a spatial system size of 5000, recording data at 
spatial points at gaps 
of 10. 

We focus on the parameter region labelled `C' or `Complex' in \cite{rien1,rien2}, 
where the order-parameter dynamics shows temporal chaos. 
We begin by verifying that the inclusion of spatial variation does not alter 
the phase diagram radically. 
To this end, we construct local phase portraits 
(plotting different components of the order parameter against each other) 
as a function of the parameters entering into the model, the shear rate 
$\dot{\gamma}$ and the tumbling parameter $\lambda_{k}$. Shown 
in Fig. \ref{phport} are the local phase portraits for a particular 
point $x_0$ for various shear-rates $\dot{\gamma}$, obtained by holding the 
value of the tumbling parameter fixed at $\lambda_{k} = 1.25$. We have  
checked that the character of the phase portrait (space-filling or otherwise) 
remains intact upon going from one space point to another though there is no 
phase coherence between two such portraits. A closed curve corresponding to a limit cycle 
is seen in the tumbling region of parameter space (denoted by `T') in the 
figure, while in the `C' region of the phase space it is space filling. 
When one goes away from the `C' region of the phase space to the region 
where the director aligns with the flow the points reduce to those on a line 
and eventually in the aligning regime where the director has already aligned 
with the flow it is represented by a point. This assures us that the local 
dynamics in the spatially extended case is similar to that of the ODEs of 
\cite{rien1,rien2}.  

\begin{figure}
\includegraphics[width=8.5cm,height=5.2cm]{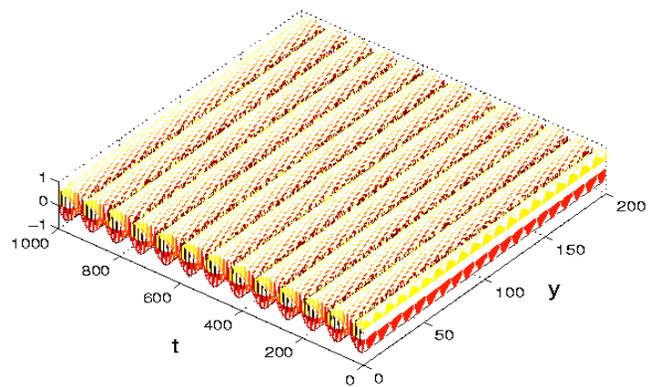}
\caption{\label{TumbShearstress}Space-time behavior of the shear 
stress in the tumbling regime,
$\dot{\gamma}=5.0$ and $\lambda_k=0.9$. Slice taken from a system of size 5000.} 
\end{figure}

\begin{figure}
\includegraphics[width=8.5cm,height=5.2cm]{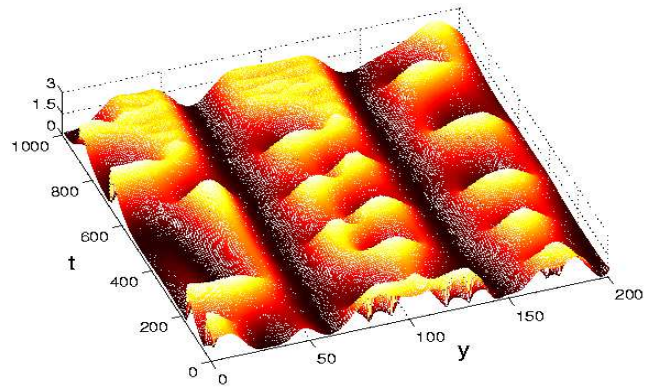}
\caption{\label{CtoAShearstress}Space-time behavior of the shear stress in the
chaotic to aligning regime, $\dot{\gamma}=4.1$ and $\lambda_k$=1.25.
Slice taken from a system of size 5000.}
\end{figure}

The contribution of the alignment tensor to the deviatoric stress is 
\cite{rien2,forster,olmstedgoldbart}  
$\sigop  \propto \alpha_0 \G - \alpha_1(\Q \cdot \G)_{ST}$ 
where $\G$, defined in (\ref{molfield}), is the nematic molecular field, 
and the total deviatoric stress is $\sigop$ plus the bare viscous stress which is 
a constant within the passive convection approximation. We 
therefore look at $\sigop$ alone.  
The spatiotemporal nature of the chaos 
\cite{ashwin} in this system is seen strikingly, 
if qualitatively, in the space-time plots of the shear stress 
(the $xy$ component of $\sigop$), refered to as $\Sigma$ henceforth, 
in Fig. \ref{ChShearstress}.  Two other figures 
are instructive, showing the periodic tumbling regime (Fig. \ref{TumbShearstress}) 
and a transition 
from spatiotemporal chaos to stable flow alignment (Fig. \ref{CtoAShearstress}). 
The parameter 
values at which these are seen, furthermore, correspond well with the phase 
diagram of \cite{rien1,rien2}. Our method therefore reveals nicely the 
rich spatiotemporal structure of the phase diagram of this system. In 
Fig. \ref{CtoAShearstress}, several distinct events, such as the persistence, movement, 
and abrupt disappearance of shear bands can be seen, whereas in Fig. 
\ref{TumbShearstress} the structure is simply periodic. 
We find that the typical length scale at which banding occurs is 
a fraction of the system size. 
In order to characterize better the 
disorderly structure of the shear bands in Fig. \ref{ChShearstress}, we look 
at the distribution of band sizes or spatial ``stress drops'' as follows. 
At a given time (say $t_i$), we define a threshold 
$\Sigma_0 = 0.8$, a little above the global mean 
$\langle \Sigma(y,t) \rangle_{y,t}$, 
and map the spatial configuration to a space-time array of $\pm 1$:  
$\tilde{\Sigma} = \mbox{sgn}(\Sigma-\Sigma_0)$. 
Fig. \ref{banddistrib} shows the histogram of the spatial length of intervals 
corresponding to the $+ $state. 
We have considered configurations extending over 2500 spatial points, and the
statistics is summed over configurations sampled at 5000 times 
(i.e. $i=1,5000$). 
We see that the 
distribution of band lengths in the spatiotemporally chaotic 
regime is fairly broad and roughly exponential in shape, and the 
time-averaged spatial correlation function of the stress also decays 
on a length scale of about 5 bands.    

The spatially averaged 
shear stress shows irregular oscillations in time, but analysis of this signal based 
on the 
techniques of \cite{tisean} does not give reliable results for estimating the correlation 
dimension $\nu$ mainly because the chaos that we observe is quite high dimensional 
(embedding dimension \cite{tisean,kennel} $m \ge 10$). A very long data train is then  
required for the analysis of the spatially averaged time series to yield a correct 
value of the correlation dimensions. We however establish conclusively 
that the oscillatory behavior of the stress and the first normal stress 
difference shows signatures of spatiotemporal chaos. 
Instead of trying to implement the correlation-dimension method for our 
spatially extended problem, we study the Lyapunov spectrum \cite{tisean,sanosawada}. 
Further, instead of studying systems of ever-increasing size, we 
look at subsystems of size $N_s$ in a given large system of size $N$, 
i.e., at space points $j$ in an interval  
$i_0 < j < i_0 + N_s -1$ where $i_0$ is an arbitrary reference point. 
For spatiotemporal chaos we expect to find that the number of 
positive Lyapunov exponents grows systematically with $N_s$, which is precisely what 
Fig. \ref{Npositivelambda} shows. For both figures in Fig. \ref{Npositivelambda} 
, we carry out the procedure with two different 
reference points $i_0$ and find essentially the same curves.  
Furthermore, it has been reported in many studies of spatiotemporally chaotic systems 
\cite{Carr} that when calculating the subsystem 
Lyapunov spectrum for increasing 
subsystem size $N_s$, one finds that the
Lyapunov exponents of two consecutive sizes are interleaved, i.e. the $i$th 
Lyapunov exponent $\lambda_i$ for the sub-system of size $N_s$ lies between the 
$i$th and ($i+1$)th Lyapunov exponent of the subsystem of size
$N_s + 1$. A direct consequence of this 
property is that with increasing subsystem size $N_s$, the largest Lyapunov 
exponent will also increase. This is clearly seen in Fig. \ref{Npositivelambda} (b). 


In summary, we have shown that the stress field in the equations of sheared, 
nonlinear nematodynamics,  
in a subrange of the parameter space where flow alignment 
does not obtain, shows spatiotemporal chaos.  
Our numerical study of this model, which extends that 
by \cite{grosso,rien1,rien2} for the purely temporal behavior, shows irregular, 
dynamic shear 
banding, and finds that the number of positive Lyapunov exponents as well 
as the largest of these grow systematically with system size. 
Since we do not know what parameter values in our model correspond to 
the experiments, a detailed comparison to data is not possible at this stage.  
However, it seems likely to us that the nonlinear relaxation of the order parameter,   
together with the coupling of nematic order parameter to flow, are the  
key ingredients for rheological chaos in a variety of problems.  
Related issues such as the effect of two- and three-dimensional variation, 
hydrodynamic flow beyond passive advection, the routes 
to spatiotemporal chaos and, of course, the role of  
other degrees of freedom such as variable micelle 
length and lifetime \cite{catesolmsted03} remain to  
be explored. Meanwhile, we look forward to quantitative 
experimental studies of spatiotemporal rheochaos \cite{cladisvansaarloos}.  

We thank G. Ananthakrishna and R. Pandit for very useful discussions, 
and SERC, IISc for computational facilities. MD acknowledges support 
from CSIR, India, and CD and SR from DST, India through the Centre for 
Condensed Matter Theory.

\end{document}